\pdfoutput=1 
\documentclass[prl,twocolumn,preprintnumbers,superscriptaddress,showpacs]{revtex4-1}

\usepackage{graphicx}
\usepackage{bm}
\usepackage{amsmath}
\usepackage{amssymb}
\usepackage{slashed}
\usepackage{natbib}
\usepackage[colorlinks,pdfstartview=FitH]{hyperref}
\hypersetup{linkcolor=blue,citecolor=blue,filecolor=black,urlcolor=blue}
\newcommand\sect[1]{\emph{#1}}

\usepackage{lipsum}

\def \be {\begin{equation} }
\def \ee {\end{equation}}

\def \bem {\begin{multline}}
\def \eem {\end{multline}}

\def \bes {\begin{subequations} }
\def \ees {\end{subequations}}

\def \pd {\partial}

\def \a {\alpha}
\def \b {\beta}

\def \d {\delta}
\def \e {\epsilon}

\def \g {\gamma}
\def \h {\eta}

\def \o {\omega}
\def \p {\pi}
\def \ps {\psi}

\def \r {\rho}
\def \l {\lambda}
\def \m {\mu}
\def \n {\nu}
\def \s {\sigma}
\def \t {\tau}

\def \x {\xi}
\def \z {\zeta}
\def \f {\phi}

\def \D {\Delta}
\def \G {\Gamma}
\def \O {\Omega}
\def \P {\Pi}

\def \<{\langle}
\def \>{\rangle}
\def \+{\dagger}
\def \({\left(}
\def \){\right)}
\def \[{\left[}
\def \]{\right]}

\def \tD {\widetilde{D}}

\usepackage[normalem]{ulem}

\begin{document}

\author{Shu~Lin}
\affiliation{School of Physics and Astronomy, Sun Yat-Sen University, Zhuhai 519082, China}
\affiliation{Guangdong Provincial Key Laboratory of Quantum Metrology and Sensing, Sun Yat-Sen University, Zhuhai 519082, China}

\title{Kinetic theories: from curved space to flat space}
\date{\today}

\begin{abstract}
	We generalize the equivalence between off-equilibrium state and gravitational perturbation of equilibrium state from dynamics of macroscopic quantities to that of microscopic particles. We also generalize the equivalence to incorporate off-equilibrium state with vorticity by torsional perturbation to equilibrium state. The equivalence is achieved by mapping kinetic theories of spinless and spinning particles in torsional curved space to flat space through suitable choice of inertial frame that eliminates geodesic forces on particles. The equivalence has been shown for hydrodynamic and elastic regimes. In the latter case, we predict spin polarization induced by time-variation of shear strain in elastic materials. We also provide an order-of-magnitude estimate for the polarization in Dirac semi-metal.
\end{abstract}
\maketitle 

\sect{\bf Introduction}%

It has been commonly believed that metric perturbation to equilibrium state can serve as a proxy for off-equilibrium state in flat space. A well-known example is that temperature variation in an off-equilibrium state can be simulated by gravitoelectric potential \cite{Luttinger:1964zz}. The underlying logic is that energy density couples to the variations of temperature and gravitoelectric potential in the same way. The logic has been extended to fluid motion \cite{Nomura:2011hn,Bradlyn:2014wla}, in which the momentum density couples to fluid velocity and gravitomagnetic potential in the same way. The logic has been implemented in deriving hydrodynamics from gravitational background \cite{Banerjee:2012iz,Jensen:2012jh}. More recently, it has been show that the gravitational background emerges naturally in time evolution of quantum fields from Gibbs ensemble method \cite{Hayata:2015lga,Hongo:2016mqm}.

While the equivalence seems natural for dynamics of macroscopic quantities such as energy and momentum densities, one may ask if the equivalence holds for more general quantities. A typical case is dynamics of microscopic particles in an off-equilibrium state, which is of wide phenomenological interests in relativistic heavy ion collisions \cite{Becattini:2020ngo}, astrophysics \cite{Chamel:2008ca,Kamada:2022nyt} and condensed matter systems \cite{Gorbar:2021ebc}. An obvious obstruction to the equivalence arises due to momentum of particle: on one hand, momenta in curved space and flat space differ; on the other hand, geodesic force changes momentum in curved space with no counterpart in flat space. Nevertheless we will show that the equivalence for dynamics of microscopic particles can still be established with a proper choice of inertial frame, which links momenta in the curved and flat spaces and simultaneously eliminate the geodesic force. Interestingly, the chosen frame has vanishing spin connection, extending the equivalence to spinning particle innocuously. The mechanism will be shown to hold in both hydrodynamic and non-hydrodynamic regimes.

Throughout this work, we consider weakly interacting energetic particles in a near equilibrium state, for which kinetic description and gradient expansion applies. We use the most minus signature. Greek and Latin indices correspond to curved and flat indices respectively. Symmetrization and anti-symmetrization are defined as $\{X,Y\}=\frac{1}{2}(XY+YX)$ and $\[X,Y\]=\frac{1}{2}(XY-YX)$ respectively.

\sect{\bf Kinetic theory in hydrodynamic background}

We start with spinless particles in curved space, whose dynamics is governed by the Boltzmann equation \cite{Stewart1971,Fonarev:1993ht}
\begin{align}\label{Boltzmann}
p^{\m} D_\m f(x,p)=-C[f],
\end{align}
with $D_\m=\nabla_\m+\G_{\n\m}^\l p_\l\pd_p^\n$ being the horizontal lift of the covariant derivative $\nabla_\m$ and $\pd_p^\n\equiv\frac{\pd}{\pd p_\n}$. We have chosen $p_\m$ and $x^\m$ as momentum and coordinate of particle respectively. $f$ is the distribution function and $\nabla_\m$ reduces to $\pd_\m^x\equiv\frac{\pd}{\pd x^\m}$. The Boltzmann equation is valid provided that separation of scale $p\gg\pd_x$ holds. \eqref{Boltzmann} is derived for torsion free space \cite{Stewart1971,Fonarev:1993ht}. For later use, we also derive its generalization to torsional space in the supplemental materials. The resulting equation in the collisionless limit is given by
\begin{align}\label{Boltzmann_torsion}
p^{\m} \(D_\m+T^\l_{\m\n}p_\l\pd_p^\n\) W(x,p)=0.
\end{align}
$W(x,p)=\int d^4y\sqrt{-g(x)}e^{-ip\cdot y}\langle \f_-(x,y)\f_+(x,y)\rangle$ is the Wigner function for scalar particle with $\f_\pm(x,y)\equiv e^{\pm y\cdot D^y/2}\f(x)$ and $D^y_\m=\nabla_\m-\G_{\n\m}^\l y^\n\pd_\l^y$ is the Fourier transform of $D_\m$ introduced such that $W(x,p)$ is diffeomorphism invariant. $W(x,p)$ is related to the distribution function as $W(x,p)=\d(p^2-m^2)f(x,p)$. Although scalar doesn't couple to torsion, its momentum variable defined in the cotangent space still senses the presence of torsion. Indeed, torsion term appears as a correction to the horizontal lift. 

In order to simulate off-equilibrium state in flat space, we requires $f(x,p)$ to be local equilibrium distribution of the curved space $f(x,p)=f(p^\m u_\m/T)$ with $u_\m=(\sqrt{g_{00}},0,0,0)$ and $T=T_0/\sqrt{g_{00}}$ being local fluid velocity and temperature \cite{Banerjee:2012iz}. Using that $D_\m\d(p^2-m^2)=0$, we can pull $\d(p^2-m^2)$ to the front. For the torsional term, we can use integration by part as $T^\l_{.\m\n}p_\l\pd_p^\n(\d(p^2-m^2) f(x,p))=-\pd_p^\n(T^\l_{.\m\n}p_\l)(\d(p^2-m^2) f(x,p))=-T_\m \d(p^2-m^2) f(x,p)$ with $T_\m\equiv T^\l_{.\m\l}$. We then simplify \eqref{Boltzmann_torsion} as
\begin{align}\label{disp_pull}
\d(p^2-m^2) p^\m (D_\m-T_\m)f(x,p)=0.
\end{align}
To reduce to the collisionless limit of \eqref{Boltzmann}, we impose the constraint $T_\m=0$. As will be clear soon, the constraint follows naturally from hydrodynamic equations.

Now we map this solution to off-equilibrium state in flat space by switching to flat space counterpart of $p^\m$ and $\b_\m\equiv u_\m/T=(g_{00},0,0,0)/T_0$. This amounts to choosing a local inertial frame specified by vielbein $e^\m_a$ as
\begin{align}
p^\m=e^\m_a p^a,\quad \b_\m=e_\m^a \b_a.
\end{align}
We wish to describe off-equilibrium state in hydrodynamic regime characterized by $\b_a=(1,-u^i)/(T_0+\d T)$ with $\d T$ and $u^i$ being variation of temperature and fluid velocity respectively. Using $g_{\m\n}=e_\m^a e_\n^b\h_{ab}$ and the explicit expressions of $\b_\m$ and $\b_a$, we find the following vielbein and metric perturbations to first order in these fields
\begin{align}\label{vielbein_hydro}
e_0^{\hat{0}}=1-\frac{\d T}{T_0},\; e_i^{\hat{0}}=u^i,\quad g_{00}=1-\frac{2\d T}{T},\; g_{0i}=u^i. 
\end{align}
The metric perturbations are the widely-used one for simulating hydrodynamic quantities in the torsion free case.
Since we have already simulate all hydrodynamic fields, we may set the remaining components of vielbein as $e_0^{\hat{i}}=0$ and $e_i^{\hat{j}}=\d_{ij}$.

To reduce \eqref{Boltzmann} to its flat space counterpart, we still need to eliminate the geodesic force, which has no counterpart in flat space. The crucial observation is the following: non-uniform frame choice mapping $p^\m$ to $p^a$ leads to an additional frame force on $p^a$. We can use the freedom of frame choice to cancel the geodesic force. 
To see this, we rewrite the collisionless limit of \eqref{Boltzmann} as
\begin{align}\label{forces}
p^\m p_\l\(e_a^\l\pd_\m\b^a+\b^a\pd_\m e_a^\l+\G_{\n\m}^\l\b^a e_a^\n\)\frac{\pd f}{\pd (\b\cdot p)}=0.
\end{align}
The second and third terms correspond to frame force and geodesic force respectively. The condition for the cancellation of geodesic force by the frame force is given by
\begin{align}\label{force_cancel}
\pd_\m e_a^\l+\G_{\n\m}^\l e_a^\n=0.
\end{align}
Using vielbein postulate $0=\nabla_\m e_a^\l=\pd_\m e_a^\l+\G_{\n\m}^\l e_a^\n+\o_{\m,ab}e^{\l b}$, we easily find a simple solution $\o_{\m,ab}=0$. 
The remaining first term can be written using $p^\m\pd_\m= p^b\pd_b$, $p^\n e_\n^a= p^a$ as
\begin{align}\label{reduction_hydro}
p^b p_a\pd_b \b^a\frac{\pd f}{\pd(\b\cdot p)}=0,
\end{align}
which is the collisionless Boltzmann equation in off-equilibrium state in flat space.
With spin connection vanished and vielbein fixed in \eqref{vielbein_hydro}, the curved space is uniquely determined and we can obtain the curvature and torsion from Maurer-Cartan equation as
\begin{align}\label{cartan}
T^a_{.\m\n}=\pd_\m e_\n^a-\pd_\n e_\m^a,\quad
R^a_{b\m\n}=0.
\end{align}
The introduction of torsion is essential for describing vorticity. \eqref{cartan} gives the following relationship between vorticity and torsion tensor.
\begin{align}\label{torsion_vorticity}
T_{.ij}^{\hat{0}}=\pd_i u^j-\pd_j u^i=2\e^{ijk}\O^k,
\end{align}
with $\O^k$ being the vorticity.

We are ready to verify the constraint $T_\m=0$. With the hydrodynamic fields encoded in \eqref{vielbein_hydro} only, we find that to linear order in the hydrodynamic fields $T_0=0$ is trivially satisfied. $T_i=0$ is spelled out as follows
\begin{align}\label{torsion-constraint}
T_i=\partial_i e_0^{\hat{0}}-\partial_0 e_i^{\hat{0}}=-\frac{\pd_i\d T}{T_0}-\pd_0 u^i=0.
\end{align}
We show that it follows from ideal hydrodynamic equations $\pd_\n T^{\m\n}=0$. With $T^{\m\n}=(\e+p)u^\m u^\n-p \h^{\m\n}$, we have from $\m=i$ component that
\begin{align}
(\e+p)\pd_0 u^i+\pd_i\d p=0.
\end{align}
This leads to the constraint $\pd_0u^i=-\frac{\pd_i\d p}{\e+p}=-\frac{\pd_i\d T}{T}$, upon using the thermodynamic relation $d p=s d T=\frac{(\e+p)d T}{T}$.

With $T_\m=0$ and \eqref{reduction_hydro}, we have shown that the left hand side (LHS) of the Boltzmann equation in curved space maps to the flat space counterpart with the chosen inertial frame.
%
Now we show how the mapping works for the collision term on the right hand side (RHS). 
The generic form of collision term in curved space is given by \cite{Stewart1971}
\begin{align}\label{collision}
&C[f]=\int d\P_{p'}d\P_{k}d\P_{k'}|{\cal M}|^2(2\p)^4\sqrt{-g}\d^4(p+p'-k-k')\nonumber\\
&\(f_pf_{p'}(1-f_k)(1-f_{k'})-f_kf_{k'}(1-f_p)(1-f_{p'})\),
\end{align}
with $d\P_{p'}=\frac{1}{\sqrt{-g}}\frac{d^4p'}{(2\p)^3}\d(p'{}^2-m^2)$. All momenta appearing in \eqref{collision} are covariant. The form of $C[f]$ can be derived from Kadanoff-Baym equation \cite{Hohenegger:2008zk}. The key observation is that $C[f]$ senses the effect of curved space only through propagators of scalar rather than vertices. The propagator at the order we keep is dictated by the dispersion relation and distribution function, which are identical to the flat space counterpart. The phase space $d\P_{p'}d\P_{k}d\P_{k'}\sqrt{-g}\d^4(p+p'-k-k')$ rewritten in terms of flat space momenta with $p^a=e^a_\m p^\m$ simply amounts to the replacement $p^\m\to p^a$, $\sqrt{-g}\to 1$. It follows that there is no curved space effect in the collision term. Thus we conclude that the collision term of Boltzmann equation in curved and flat spaces are also identified.

\sect{\bf Kinetic theory in elastic background}

The previous discussions show how dynamics of microscopic particles in a flat space off-equilibrium state in hydrodynamic regime is equivalent to counterpart in curved space equilibrium state. The mechanism can be extended to non-hydrodynamic regime as well. An obvious extension is to dynamics of particles in an elastic matter. Off-equilibrium state in this case corresponds to collective deformation of lattice in the elastic matter. The geometric description of elastic matter is based on a mapping between physical space and matter space \cite{DeWitt:1962cg,Carter1972} \footnote{Notably the geometric picture has also been generalized to the hydrodynamic regime in \cite{Crossley:2015evo}}.
\begin{align}\label{lie}
x^0=\z^0+\x^0(x),\quad x^i=\z^i+\x^i(x),
\end{align}
with $x^\m$ and $\z^a$ being the coordinates of physical space and matter space. $\z^0$ and $\z^i$ correspond to internal time and equilibrium position of matter elements. $\x^i$ is the displacement field and $\x^0$ parameterizes the deviation of internal time from physical time. The mapping characterized by the vielbein $e_\m^a=\frac{\pd \z^a}{\pd x^\m}$ determines a curved physical space from a flat matter space. The physics is expected to be invariant under reparameterization of internal time, thus we may fix $\z^0=x^0$. The corresponding nonvanishing vielbein components given by
\begin{align}
e_0^{\hat{0}}=1,\quad e_0^{\hat{i}}=-\pd_0\x^i,\quad e_j^{\hat{i}}=\d_{ij}-\pd_j\x^i.
\end{align}

We now consider microscopic particles in the deformed matter background. We shall assume slow-varying displacement field $\x^i(x^0,x^i)$ with the separation of scale $p\gg \pd_x \x/\x$, which is expected to hold for phonon excitation. In this case, the Boltzmann equation \eqref{Boltzmann_torsion} can still be used. The local equilibrium distribution in this case is given by $f(\frac{p^\m U_\m}{T})$, which is formally the same as in the hydrodynamic background. There are two differences: i) here $U^\m=\frac{dx^\m}{d\t}$ is the velocity of matter elements rather than fluid velocity, with $\t$ being the proper time such that $U^2=1$. A slow-varying matter velocity field corresponds to phonon excitation; ii) the displacement field is treated as arbitrary and we don't impose equation of motion like in the hydrodynamic case. Only the microscopic particles surfing on the phonon background are assumed in local equilibrium with constant temperature $T$.

Similar to the hydrodynamic case, we map the curved momentum and equilibrium velocity to their flat space counterparts by vielbein as $p^a=e_\m^a p^\m$, $U_a=e^\m_a U_\m$. The condition for cancellation of frame force against geodesic force is analogous to \eqref{force_cancel}:
\begin{align}\label{frame_cond}
U_a\(\pd_\m e_a^\l-\G_{\n\m}^\l e_a^\n\)=0.
\end{align}
This is again solved by vanishing spin connection. It follows that \eqref{cartan} is also applicable for the elastic case. It remains to verify the constraint $T_\m=0$ needed for the mapping to the flat space Boltzmann equation. In fact, we readily verify the torsion tensor vanishes identically as \footnote{Nonvanishing torsion is allowed in the presence of dislocation in matter \cite{Katanaev:1992kh}. We leave the interesting possibility for future studies.}
\begin{align}
&T^{\hat{0}}_{.\m\n}=0,\nonumber\\
&T^{\hat{i}}_{.\m\n}=\pd_\m e_\n^{\hat{i}}-\pd_\n e_\m^{\hat{i}}=-\pd_\m\pd_\n \x^{\hat{i}}+\pd_\n\pd_\m \x^{\hat{i}}=0.
\end{align}
Similar to \eqref{reduction_hydro}, we arrive at the following Boltzmann equation
\begin{align}\label{reduction_elastic}
p^b p^a\pd_b U_a\frac{\pd f}{\pd (U\cdot p)}=-C[f].
\end{align}
The discussions on mapping of the collision term in the hydrodynamic case can be carried over to the elastic case. Thus we have shown that the equivalence between off-equilibrium state and gravitational perturbation of equilibrium state for dynamics of microscopic particles holds in elastic regime as well.


\sect{\bf Quantum kinetic theory for spinning particles}

So far we have restricted ourselves to spinless particles. A natural question to ask is whether the mechanism extends to spinning particles? A complication compared to the spinless case is that spin also senses the presence of torsion in addition to momentum. Indeed, spinning particles couples to torsion implicitly through spin connection. We shall show below that equivalence continue to work as the chosen local inertial frame has vanishing spin connection.

We first work in the collisionless limit and address collisional effect at the end. The quantum kinetic equation for massive spin one half particle in torsion free space has been derived in \cite{Liu:2020flb,Liu:2018xip}. In the supplementary materials, we generalize to torsional case up to $O(\pd_x)$, with the resulting equations given by
\begin{align}\label{qkt}
\(\g^\m p_\m-m+\frac{i}{2}\g^\m (D_\m+T^\l_{\m\n}p_\l\pd_p^\n)\) W(x,p)=0.
\end{align}
$W(x,p)$ is the Wigner function for Dirac particles defined as \begin{align}\label{Wigner}
&W_{\a\b}(x,p)=\int d^4y\sqrt{-g(x)}e^{-ip\cdot y}\times\nonumber\\
&
\([e^{y\cdot D^y/2}\ps(x)]^\dagger\g^{\hat{0}}\)_\b \(e^{-y\cdot D^y/2}\ps(x)\)_\a,
\end{align}
$\a,\b$ are Dirac indices and $D^y_\m=\nabla_\m-\G_{\n\m}^\l y^\n\pd_\l^y$ is formally the same as the scalar counterpart except $\nabla_\m$ acting on spinor contains spin connection terms. Note that the Dirac structure of $W$ is defined in local inertial frame, as can be seen from the presence of $\g^{\hat{0}}$ in \eqref{Wigner}. This suits our purpose of describing spinning particle dynamics in flat space. We decompose the Wigner function as
\begin{align}\label{clifford}
W=S+i\g^5P+V_a\g^a+A_a\g^5\g^a+\frac{1}{2}\s^{ab}T_{ab}.
\end{align}
The coefficients of the decomposition correspond to densities of scalar, pseudoscalar, vector, axial vector and tensor respectively. In particular, $A_a$ is related to spin polarization in local inertial frame. Equation of motion for each components can be derived by using the hermitian property of the Wigner function: $W(x,p)=\g^{\hat{0}}W(x,p)^\dagger\g^{\hat{0}}$, which leads to real densities in \eqref{clifford}. Using that $D_\m W=\pd_\m W+\frac{1}{4}\o_{\m,ab}\[\g^{ab},W\]+\G_{\n\m}^\l p_\l\pd_p^\n W$, we can decompose \eqref{qkt} and separate the real and imaginary parts of the resulting coefficients to have
\begin{subequations}\label{eom}
	\begin{align}
	&p_a V^a-mS=0,\label{eq11}\\
	&\frac{1}{2}\tD_a V^a=0,\label{eq12}\\
	&p_a A^a=0,\label{eq13}\\
	&\frac{1}{2}\tD_a A^a+mP=0,\label{eq14}\\
	&p^c S-\tD_a\frac{T^{ac}}{2}-mV^c=0,\label{eq15}\\
	&\frac{1}{2}\tD_c S+p^a T_{ac}=0,\label{eq16}\\
	&p^f P-\frac{1}{2}\tD_a\frac{T_{cd}}{2}\e^{acdf}=0,\label{eq17}\\
	&\frac{1}{2}\tD_b P+p^a\frac{T^{cd}}{2}\e_{acdb}-m A_b=0,\label{eq18}\\
	&p^{[b}V^{f]}+\frac{1}{4}\tD_c A_g\e^{cgdf}=0,\label{eq19}\\
	&\frac{1}{2}\tD^{[d}V^{f]}-\frac{1}{2}p_cA_g\e^{cgdf}-\frac{m}{2}T^{df}=0.\label{eq20}
	\end{align}
\end{subequations}
Details of the derivation are reserved in supplementary materials. Here $\tD_a=e_a^\m\(\nabla_\m+\G_{\n\m}^\l p_\l\pd_p^\n+T^\l_{.\m\n}p_\l\pd_p^\n\)$ and the covariant derivative acts on flat indices as $\nabla_\m A^a=\pd_\m A^a+\o_{\m}^a{}_b A^b$.

Now we comment on how the set of equations can be mapped to their flat space counterpart. In the absence of torsion, $W\propto\d(p^2-m^2)$ up to $O(\pd_x)$ \cite{Liu:2020flb,Liu:2018xip}. The presence of torsion doesn't alter this at this order for the following reason: generic $T^\l_{.\m\n}$ can be decomposed as \cite{Shapiro:2001rz}
\begin{align}\label{torsion_decomp}
T_{\a\b\m}=\frac{1}{3}\(T_\b g_{\a\m}-T_\m g_{\a\b}\)-\frac{1}{6}\e_{\a\b\m\n}S^\n+q_{\a\b\m},
\end{align}
with $T_\m$, $S_\m$ and $q_{\a\b\m}$ being vector, axial and tensor components of torsion. At $O(\pd_x)$, we can't form corrections to $p^2$, which are even under both time-reversal symmetry and parity: possible forms are $p^\m T_\m$, $p^\m S_\m$ and $p^\b g^{\a\m}q_{\a\b\m}$. The first two are odd under time-reversal and parity respectively. The last one vanishes identically. Thus we deduce $W\propto\d(p^2-m^2)$ remains correct in the presence of torsion. We can then use similar logic above \eqref{disp_pull} to rewrite $\tD_a=e_a^\m\(\nabla_\m+\G_{\n\m}^\l p_\l\pd_p^\n-T_\m\)$. We will use the same curved spaces and choice of frame for scalar particles to simulate off-equilibrium dynamics for spinning particles. In the hydrodynamic case, we have verified $T_\m=0$ so that $\tD_a=D_a$. Components of Wigner function are expected to be functions of $p^a$, $\b^a$ and $f(p\cdot\b)$ (including its derivatives) for the hydrodynamic case. With vanishing spin connection, we readily verify that $D_a p^b=0$, $D_a\b^b=\pd_a\b^b$ and $D_a f(p\cdot\b)=p_b\pd_a\b^b(\pd f/\pd(p\cdot\b))$, i.e. all curved space effects vanish, thus we have mapped the quantum kinetic equations for spinning particles in curved space \eqref{eom} to the flat space counterpart. The conclusion obviously holds for elastic case from the substitution $\b\to U/T$ with constant $T$.

Now we solve \eqref{eom} to obtain spin polarization of particles in off-equilibrium state in both hydrodynamic and elastic regimes. For this purpose, we can assume the following scaling for different components: $V^a,\,S\sim O(1)$, $A^a,\,T^{ab}\sim O(\pd_x)$ and $P\sim O(\pd_x^2)$ \cite{Yang:2020hri}. With the scaling above and power counting $p_a, e_a^\m\sim O(1)$, $D_a\sim O(\pd_x)$, we may keep only equations up to $O(\pd_x)$, reducing \eqref{eom} to the following subset
\begin{subequations}
	\begin{align}
	&p_a V^a-mS=0,\label{eq11r}\\
	&\frac{1}{2}D_a V^a=0,\label{eq12r}\\
	&p_a A^a=0,\label{eq13r}\\
	&p^c S-D_a\frac{T^{ac}}{2}-mV^c=0,\label{eq15r}\\
	&\frac{1}{2}D_c S+p^a T_{ac}=0,\label{eq16r}\\
	&p_a\frac{T_{cd}}{2}\e^{acdb}-m A^b=0,\label{eq18r}\\
	&\frac{1}{2}D^{[d}V^{f]}-\frac{1}{2}p_cA_g\e^{cgdf}-\frac{m}{2}T^{df}=0,\label{eq20r}
	\end{align}
\end{subequations}
\eqref{eq11r} and \eqref{eq15r} can be solved by (correction from the term $D_aT^{ac}$ is of order $O(\pd_x^2)$)
\begin{align}\label{VS}
&V^a=p^a\d(p^2-m^2)f(x,p),\nonumber\\
&S=m\d(p^2-m^2)f(x,p).
\end{align}
Plugging \eqref{VS} into \eqref{eq12r}, and using $D_a p^a=0$, we obtain the Boltzmann equation \eqref{Boltzmann}. As in the spinless case, we choose the local equilibrium distribution. Focusing on particles with positive energy, we have $f(x,p)={2\p}{(e^{p\cdot\b}+1)^{-1}}$ in the hydrodynamic case and $f(x,p)={2\p}{(e^{p\cdot U/T}+1)^{-1}}$ in the elastic case.

Now we solve for components at $O(\pd_x)$, including $A^a$ and $T^{ab}$. We will solve equations \eqref{eq12r}, \eqref{eq13r}, \eqref{eq16r}, \eqref{eq18r} and \eqref{eq20r} \footnote{We do not consider \eqref{eq19} for the following reason: \eqref{VS} only gives the solution at $O(1)$. It can be seen from \eqref{eq15} that there exist corrections at $O(\pd_x^2)$. \eqref{eq19} is only nonvanishing from $O(\pd_x^2)$, which we do not keep.}. In fact, we can easily show only \eqref{eq18r} and \eqref{eq20r} are independent by using Boltzmann equation and the dispersion relation enforced by $\d(p^2-m^2)$. 
The independent equations can be solved by
\begin{align}\label{A_sol}
&A^a=-\e^{abcd}\frac{u_b p_c D_d f}{2p\cdot u}\d(p^2-m^2),\nonumber\\
&T_{ab}=\frac{m}{2p\cdot u}\(u_a D_b f-u_b D_a f\)\d(p^2-m^2),
\end{align}
with $u_b$ being the fluid velocity in the hydrodynamic case and $u_b\to U_b$ for the elastic case. The quantity of our interest $A^a$ can then be written in terms of hydrodynamic gradients as
\begin{align}\label{A_hydro}
&A^a=-\frac{\e^{abcd}u_bp_c}{2p\cdot u}\(\frac{p^e\pd_d u_e}{T}-\frac{p\cdot u\pd_d T}{T^2}\)f'\d(p^2-m^2)\nonumber\\
&=-\frac{\e^{abcd}u_bp_c}{2p\cdot u}\(\frac{p^e\s_{de}}{T}+\frac{p^e\e_{defg}u^f\O^g}{T}-\frac{p\cdot u\pd_d T}{T^2}\)f'\nonumber\\
&\times\d(p^2-m^2),
\end{align}
with $f'=\frac{\pd f}{\pd (p\cdot \b)}$. To arrive at the second line of \eqref{A_hydro}, we note that $\pd_d u_e$ can be replaced by the projection $\D_{da}\D_{eb}\pd^a u^b$ with the projector $\D_{ab}=-\h_{ab}+u_au_b$. The projection is further decomposed as $\D_{da}\D_{eb}\pd^a u^b=\s_{de}+\e_{defg}u^f\o^g+\frac{1}{3}\D_{de}\D_{ab}\pd^a u^b$ with $\s_{de}=\D_{a\{d}\D_{e\}b}\pd^a u^b-\frac{1}{3}\D_{de}\D_{ab}\pd^a u^b$ being the shear tensor and $\O^a=\frac{1}{2}\e^{abcd}u_b\pd_c u_d$ being vorticity pseudovector. The last bulk term in the decomposition vanishes upon contraction with prefactors. Thus the two terms in the bracket of \eqref{A_hydro} correspond to spin responses to shear and temperature gradient. It agrees with \cite{Becattini:2021suc,Hidaka:2017auj,Wang:2024lis} using field theory and \cite{Liu:2021uhn} using linear response to metric perturbation. 
Since we have eliminated the geodesic force, ad-hoc subtraction introduced in \cite{Liu:2020dxg,Liu:2021uhn} is not needed in our approach.

The solution \eqref{A_hydro} for vorticity source is not unique. The following homogeneous solution to \eqref{eq18r} and \eqref{eq20r} is allowed
\begin{align}\label{shift}
&A^a=\frac{1}{2p\cdot u}\frac{1}{T}(p^ap\cdot\O-m^2\O^a)\d(p^2-m^2)f',\nonumber\\
&T_{ab}=\frac{m}{2p\cdot u}\frac{1}{T}p^c\e_{gabc}\O^g\d(p^2-m^2)f'.
\end{align}
In the massless limit, the above solution corresponds to shift of the distribution function due to spin-vorticity coupling \cite{Chen:2015gta,Hidaka:2017auj,Gao:2018jsi}. Adding \eqref{shift} to \eqref{A_hydro}, we arrive at the following solution for the vorticity source
\begin{align}\label{A_sol2}
&A^a=-\frac{1}{2T}(\O^a p\cdot u-u^a p\cdot\O)\d(p^2-m^2)f',\nonumber\\
&T_{ab}=\frac{m}{2T}\e_{fgab}u^f\O^g\d(p^2-m^2)f'.
\end{align}
This corresponds to spin response to vorticity, in agreement with \cite{Becattini:2013fla,Becattini:2020sww}. Two clarifications are in order: i) \eqref{A_sol2} characterizes spin polarization in a local rotation as opposed to the rigid rotation extensively studied in \cite{deOliveira:1962apw,Hehl:1990nf,Duffy:2002ss,Ambrus:2014uqa,Chen:2015hfc,Mameda:2015ria,Ambrus:2015lfr,Ebihara:2016fwa,Chernodub:2017ref,Selch:2023pap,Ambrus:2023smm}. In particular, we don't need to impose an artificial boundary as each patch of fluid element has a natural boundary. Since $x$ is a coarse-grained coordinate labeling fluid elements, we can't discuss orbital angular momentum within a fluid element; ii) In the hydrodynamic case, the vorticity is simulated by axial component of torsion as $\O^a=\frac{1}{12}S^a$ from \eqref{torsion_vorticity} and \eqref{torsion_decomp}. Thus the torsion effect may arises through both the simulated vorticity and the gravitational background. The absence of of the latter effect is in line with the absence of chiral torsion effect \cite{Ferreiros:2020uda} \footnote{Here we introduce torsion as an external source, so the backreaction of spin tensor density to torsion is not present. The backreaction is suppressed by gravitational coupling.}.

Now we turn to the elastic case. The analysis in the hydrodynamic case simply carries over with the replacement $u_a\to U_a$. Making replacement in \eqref{A_hydro} and dropping derivatives on temperature, we obtain the $A^a$ component as
\begin{align}\label{A_elastic}
A^a=-\frac{\e^{abcd}u_bp_c}{2p\cdot u}\frac{p^e\Sigma_{de}}{T}f'\d(p^2-m^2).
\end{align}
where $f'=\frac{\pd f}{\pd (p\cdot U/T)}$. $\Sigma_{de}=\D^U_{a\{d}\D^U_{e\}b}\pd^a U^b-\frac{1}{3}\D^U_{de}\D^U_{ab}\pd^a U^b$ and $\D_{ab}^U=-\h_{ab}+U_aU_b$ being analogs of shear tensor and projector in the hydrodynamic case. Working to leading order in the displacement field, we have $U^a\simeq(1,\dot{\x}^i)$ and $\Sigma_{ij}=-\frac{1}{2}(\pd_i\dot{\x}^j+\pd_j\dot{\x}^i)+\frac{1}{3}\d_{ij}\pd_k\dot{\x}^k$, with dot denoting $d/dx^0$.
\eqref{A_elastic} predicts spin polarization induced by varying shear strain in elastic matter.
We also have spin polarization induced by elastic vorticity analogous to \eqref{A_sol2} with $\O^a\to\O_U^a=\frac{1}{2}\e^{abcd}U_b\pd_c U_d$, though in this case the elastic vorticity is not simulated by torsion.

We are ready to discuss the curved space effect on collision term. The only complication compared to the scalar case is the presence of Dirac structure in the Wigner function, which is fixed by the chosen local inertial frame. Similar self-energy representation of collision term for spinning particles also indicates that curved space effect enter only through propagators not vertices, with the former dictated by the collisionless Wigner function. Systematic analysis of the collision term in terms of components of Wigner function has been carried out in \cite{Yang:2020hri} in flat space, see also \cite{Lin:2022tma,Weickgenannt:2022zxs,Lin:2024zik,Wang:2024lis,Fang:2024vds} for manifestation of collision term with specific interaction. Since the collision term is local, the only effect of curved space is to replace the flat space Wigner function by the curved space counterpart. Since the curved space effect has been completely removed in the chosen local inertial frame, we conclude the collision term of quantum kinetic equation in curved and flat spaces are also identified.

As an application, we give an order-of-magnitude estimate for the polarization of electron in Dirac semi-metal induced by shear wave. Since fermion in Dirac semi-metal contains both chiralities, we use the following as proxy for polarization. 
\begin{align}
{\cal P}^i\sim\frac{A^i}{V^0}\sim \frac{p^2}{p_0^2}\frac{\Sigma}{T}\frac{f'}{f},
\end{align}
where $p_0=p\cdot U$. The numerator and denominator can be interpreted as difference and sum of contributions from two chiralities. We estimate the shear tensor by $\Sigma\sim 2\p\n\s_{\text{max}}$, with $\n$ taken to be high frequency optical phonon with $\n\simeq 8\text{THz}$ \cite{PRL}. The shear strain is taken to be the maximum breaking strain estimated by $\s_{\text{max}}\sim 0.1$ \cite{nano_letter}. The electrons are moderately degenerate at room temperature $T=300\text{K}$, so we may take $\frac{f'}{f}\sim O(1)$. The ratio $\frac{p^2}{p_0^2}\simeq v_F^2$ is estimated with the Fermi velocity $v_F\sim 10^6\text{m/s}$ \cite{apl}. Putting all factors, we obtain a polarization of order $10^{-6}$.

\sect{\bf Conclusion and Outlook}%

We have shown that the equivalence between off-equilibrium state and gravitational perturbation to equilibrium state for macroscopic quantities generalizes to dynamics of microscopic particles. It is achieved through choice of local inertial frame, which eliminates the geodesic forces on both spinless and spinning particles, mapping kinetic theories in curved space to the counterpart in flat space. The equivalence is shown in both hydrodynamic and elastic regimes. We have also generalized the existing kinetic theories to the torsional case to simulate fluid vorticity. Finally we predict spin polarization induced by elastic shear wave and estimate its order of magnitude for Dirac semi-metal.

We have assumed separation of scales to perform gradient expansion and the existence of quasi-particles to adopt kinetic description. The former assumption is crucial for distinguishing fast dynamics of particles and slow-varying gravitational background. In fact, the obtained local inertial frame is independent of the properties of particles. Thus we expect the latter assumption may be relaxed, which is also in line with the findings in \cite{Hayata:2015lga,Hongo:2016mqm}. It can be tested in models of strongly coupled systems in which gradient expansion remains valid \cite{Li:2025zbj}.

It is more interesting to extend the analysis to next order in gradient expansion, in which spacetime curvature is expected to modify the flat space dispersion as well as the collision terms. The relationship between flat space off-equilibrium state and curved space equilibrium state may be significantly modified. We leave this for future explorations.

\sect{\bf Acknowledgments}

I thank Jiayuan Tian for collaboration in related works. I also thank F.~Becattini, K.~ Mameda, X.-G.~Huang, S.~Pu and Y.~Yin for stimulating discussions. This work is in part supported by NSFC under Grant Nos 12475148 and 12075328.

\section{Derivation of kinetic equations in torsional space}

In torsional space, we have to be careful with ordering of indices in covariant derivatives. We follow the convention in \cite{Shapiro:2001rz} with $\nabla_\m V_\n=\pd_\m V_\n-\G_{\n\m}^\l V_\l$. We further impose vielbein postulate $\nabla_\m e_\n^a=0$ and metricity condition $\nabla_\m g^{\m\n}=0$. It follows that $\nabla_\m \h^{ab}=0$, which indicates that the spin connection is anti-symmetric in flat indices $\nabla_\m\h^{ab}=\o_\m^{ab}+\o_\m^{ba}=0$.

We shall first derive the quantum kinetic equation for spin $1/2$ particle, whose equation of motion is first order. The quantum generalization of distribution function is the Wigner function, which is defined as
\begin{align}\label{Wigner}
&W_{\a\b}(x,p)=\int d^4y\sqrt{-g(x)}e^{-ip\cdot y}\r_{\a\b}(x,y),\nonumber\\
&\r_{\a\b}(x,y)=\langle\overline{\ps}_\b(x,y/2)\otimes\ps_\a(x,-y/2)\rangle,
\end{align}
with $\ps(x,-y/2)=e^{-y\cdot D_y/2}\ps(x)$ and $\overline{\ps}(x,y/2)=[e^{y\cdot D_y/2}\ps(x)]^\dagger\g^{\hat{0}}$.
$y^\n$ is chosen to be the conjugate of momentum $p_\n$ and $D_\m^y=\nabla_\m-\G_{\n\m}^\l y^\n\pd_\l^y$ is the horizontal lift. The affine connection term is included in $D_\m^y$ such that $D_\m^y y^\n=0$. It eliminates action of affine connection on $y^\n$ such that comparison of phase space variables at different $x$ is meaningful. It follows immediately the horizontal lift on $p_\n$: $D_\m=\nabla_\m+\G_{\n\m}^\l p_\l\pd_p^\n$, which satisfies $D_\m p_\n=0$.

Using the short hand notation $\ps_\pm(x,y)=e^{\pm y\cdot D^y/2}\ps(x)$, we derive the following relation
\begin{align}\label{partial_psm}
&\pd_\m^y\ps_-=e^{-y\cdot D^y/2}\(e^{y\cdot D^y/2}\pd_\m^y e^{-y\cdot D^y/2}\)\ps(x)\nonumber\\
&=-e^{-y\cdot D^y/2}\(\frac{D^y_\m}{2}+\frac{1}{2}[\frac{y\cdot D^y}{2},\frac{D^y_\m}{2}]+\cdots\)\ps(x),
\end{align}
where Baker-Campbell-Hausdorff formula has been used.
Similarly
\begin{align}\label{particle_psp}
\pd_\m^y\ps_+=e^{-y\cdot D^y/2}\(\frac{D^y_\m}{2}+\frac{1}{2}[-\frac{y\cdot D^y}{2},\frac{D^y_\m}{2}]+\cdots\)\ps(x).
\end{align}
The commutators in \eqref{partial_psm} and \eqref{particle_psp} is evaluated using the following relation
\begin{align}
&[D^y_\m,D^y_\n]\ps(x)=\frac{1}{4}R_{ab\m\n}\g^{ab}\ps(x)-R^\b_{\a\m\n}y^\a\pd_\b^y\ps(x)+\nonumber\\
&T^\l_{.\m\n}D^y_\l\ps(x),
\end{align}
with $T^\l_{.\m\n}=\G_{\m\n}^\l-\G_{\n\m}^\l$. Since we aim at deriving kinetic equations up to $O(\pd_x)$, we may ignore the curvature terms in the commutator. The torsion term is counted as $O(\pd_x)$ from the torsion tensor and $D^y_\l\ps\sim O(1)$. Thus we have $[D^y_\m,D^y_\n]\ps(x)\simeq T^\l_{.\m\n}D_\l\ps(x)$ so that \eqref{partial_psm} and \eqref{particle_psp} become
\begin{align}
&\pd_\m^y\ps_-=-e^{-y\cdot D^y/2}\(\frac{D^y_\m}{2}-\frac{1}{8}y^\n T^\l_{.\m\n}D^y_\l\)\ps(x),\nonumber\\
&\pd_\m^y\ps_+=e^{y\cdot D^y/2}\(\frac{D^y_\m}{2}+\frac{1}{8}y^\n T^\l_{.\m\n}D^y_\l\)\ps(x).
\end{align}
On the other hand, we can use the similar procedure and logic as above to derive
\begin{align}\label{D_ps}
&D^y_\m\ps_-=e^{-y\cdot D^y/2}\(D^y_\m-\frac{1}{2}y^\n T^\l_{.\m\n}D^y_\l\)\ps(x),\nonumber\\
&D^y_\m\ps_+=e^{y\cdot D^y/2}\(D^y_\m+\frac{1}{2}y^\n T^\l_{.\m\n}D^y_\l\)\ps(x).
\end{align}
Now we can calculate
\begin{align}\label{partial_rho}
&\pd_\m^y\r=\overline{\ps}_+\otimes\(-e^{-y\cdot D^y/2}\(\frac{D^y_\m}{2}-\frac{1}{8}y^\n T^l_{.\m\n}D^y_\l\)\ps(x)\)+\nonumber\\
&\overline{\ps}(x)\(\frac{\overleftarrow{D}^y_\m}{2}+\frac{1}{8}y^\n T^\l_{.\m\n}\overleftarrow{D}^y_\l\) e^{y\cdot\overleftarrow{D}^y/2}\otimes\ps_-\nonumber\\
&=-\overline{\ps}_+\otimes\frac{D^y_\m}{2}\ps_--\frac{1}{8}\overline{\ps}_+\otimes e^{-y\cdot D^y/2}y^\n T^\l_{.\m\n}D^y_\l\ps(x)+\nonumber\\
&\overline{\ps}_+\frac{\overleftarrow{D}^y_\m}{2}\otimes\ps_--\frac{1}{8}\overline{\ps}(x)y^\n T^\l_{.\m\n}\overleftarrow{D}^y_\l e^{y\dot D^y/2}\otimes\ps_-\nonumber\\
&=\frac{D^y_\m}{2}\r-\overline{\ps}_+\otimes D^y_\m\ps_--\frac{1}{8}y^\n T^\l_{.\m\n}D^y_\l\r.
\end{align}
We have used \eqref{partial_psm} and \eqref{particle_psp} in the first line, and used \eqref{D_ps} in the second line respectively. In the third line, we have swapped $e^{\pm y\cdot D^y/2}$ with $D^y_\l$ as the corrections are $O(\pd_x^2)$. Furthermore, the last term can be dropped as $D^y_\l\r\sim O(\pd_x)$ \footnote{The power counting is not in contradictory with $D^y_\l\ps(x)\sim O(1)$ as $\r$ is a bispinor with $\ps_\pm$ carrying nearly opposite momenta. The momenta difference conjugate to $x$ is $O(\pd_x)$}.

\eqref{partial_rho} is an identity. To derive the kinetic equation for $\r$, we need to use the equation of motion (EOM). With the following Dirac spinor action in torsional space \cite{Shapiro:2001rz}\footnote{Signs in the original reference are adapted to Minkowski signature.}
\begin{align}
S=\int d^4x\sqrt{-g}\overline{\ps}\(\g^\m\nabla_\m-\frac{1}{2}\g^\m T_\m+im\)\ps.
\end{align}
It leads to the following EOM
\begin{align}\label{EOM_torsion}
\(\g^\m\nabla_\m-\frac{1}{2}\g^\m T_\m+im\)\ps=0.
\end{align}
Contracting $\g^\m$ with \eqref{partial_rho} from the left and using \eqref{D_ps}, we have for the $\ps$ term
\begin{align}
&\g^\m D^y_\m\ps_-=\g^\m e^{-y\cdot D^y/2}\(D^y_\m-\frac{1}{2}y^\n T^\l_{.\m\n}D^y_\l\)\ps(x)\nonumber\\
&=e^{-y\cdot D^y/2}\g^\m D^y_\m\ps(x)-\frac{1}{2}\g^\m y^\n T^\l_{.\m\n}D^y_\l\ps_-\nonumber\\
&=e^{-y\cdot D^y/2}\(\frac{1}{2}\g^\m T_\m-im\)\ps(x)-\frac{1}{2}\g^\m y^\n T^\l_{.\m\n}D^y_\l\ps_-.
\end{align}
In the second line, we have swapped $e^{-y\cdot D^y/2}$ with $\g^\m$ because $[D^y_\m,\g^\n]=0$ \footnote{This follows from $\nabla_\m\g^\n=e_a^\n\(\pd_\m\g^a+\o_\m^a{}_b\g^b+\frac{1}{4}\o_{\m cd}[\g^{cd},\g^a]\)=0$ by using $[\g^{cd},\g^a]=2\(\g^c\h^{da}-\g^d\h^{ca}\)$.}. We have also swapped $e^{-y\cdot D^y/2}$ with $D^y_\l$ as the correction is $O(\pd_x^2)$. In the third line, the EOM \eqref{EOM_torsion} has been used. Thus we obtain
\begin{align}
&\g^\m\pd_\m^y\r=\frac{1}{2}\g^\m D^y_\m\r-\overline{\ps}_+\otimes\(\frac{1}{2}\g^\m T_\m-im\)\ps_-+\nonumber\\
&\overline{\ps}_+\otimes\frac{1}{2}\g^\m y^\n T^{\l}_{.\m\n}D^y_\l\ps_-.
\end{align}
It remains to evaluate $\overline{\ps}_+\otimes D^y_\l\ps_-$ to $O(1)$. This can be done using the following relation
\begin{align}
&\overline{\ps}_+\otimes D^y_\m\ps_-+\overline{\ps}_+\overleftarrow{D}^y_\m\otimes\ps_-=D^y_\m\r\simeq 0,\nonumber\\
&\overline{\ps}_+\otimes D^y_\m\ps_--\overline{\ps}_+\overleftarrow{D}^y_\m\otimes\ps_-=-2\pd_\m^y\r.
\end{align}
The second relation follows from \eqref{partial_psm} and \eqref{particle_psp}. We readily obtain $\overline{\ps}_+\otimes D^y_\m\ps_-\simeq -\pd_\m^y\r$.
Applying the Wigner transform $\int d^4y\sqrt{-g}e^{-i p\cdot y}$ and using integration by part, we arrive at
\begin{align}
\(\g^\m p_\m-m\)W+\frac{i}{2}\g^\m\(D_\m+T^\l_{.\m\n}p_\l\pd_p^\n\)W=0.
\end{align}
We can further use the explicit form of $D_\m$ to rewrite the above as
\begin{align}
\(\g^\m p_\m-m\)W+\frac{i}{2}\g^\m\(\nabla_\m+\G^\l_{\m\n}p_\l\pd_p^\n\)W=0.
\end{align}

Now we turn to the derivation of kinetic equation for scalar. The procedure and logic is similar to the Dirac spinor case, except that the EOM is second order. We shall keep only key steps. The Wigner function is defined similarly as
\begin{align}
&W(x,p)=\int d^4y\sqrt{-g(x)}e^{-ip\cdot y}\r(x,y),\nonumber\\
&\r(x,y)=\langle\f_-(x,y)\f_+(x,y)\rangle,
\end{align}
with $\f_\pm(x,y)=e^{\pm y\cdot D^y/2}\f(x)$. The analogs of \eqref{partial_psm} and \eqref{particle_psp} are given by
\begin{align}\label{partial_phi}
&\pd_\m^y\f_-=-\(\frac{D^y_\m}{2}-\frac{1}{8}[y\cdot D^y,D^y_\m]\)\f_-,\nonumber\\
&\pd_\m^y\f_+=\(\frac{D^y_\m}{2}+\frac{1}{8}[y\cdot D^y,D^y_\m]\)\f_+.
\end{align}
Using \eqref{partial_phi}, we have
\begin{align}\label{DD_fm}
&\(\pd_\m^y+\frac{D^y_\m}{2}-\frac{1}{8}[y\dot D^y,D^y_\m]\)\(\pd_\n^y+\frac{D^y_\n}{2}-\frac{1}{8}[y\cdot D^y,D^y_\n]\)\r\nonumber\\
&=\f_-\(\pd_\m^y+\frac{D^y_\m}{2}-\frac{1}{8}[y\cdot D^y,D^y_\m]\)\(D^y_\n\f_+\)\nonumber\\
&=\f_-(D^y_\n D^y_\m\f_+)+\f_-\frac{1}{2}([D^y_\m,D^y_\n]\f_+)\nonumber\\
&=\f_-(D^y_\n D^y_\m\f_+)+\f_-\frac{1}{2}(T^\l_{.\m\n}D^y_\l\f_+).
\end{align}
Similarly
\begin{align}\label{DD_fp}
&\(\pd_\m^y-\frac{D^y_\m}{2}-\frac{1}{8}[y\dot D^y,D^y_\m]\)\(\pd_\n^y-\frac{D^y_\n}{2}-\frac{1}{8}[y\cdot D^y,D^y_\n]\)\r\nonumber\\
&=(D^y_\n D^y_\m\f_-)\f_++\frac{1}{2}T^\l_{.\m\n}(D^y_\l\f_-)\f_+.
\end{align}
Note that $D^y_\l\r\sim O(\pd_x)$. We can then drop the commutators on the left hand side (LHS) of \eqref{DD_fm} and \eqref{DD_fp}. We will further contract \eqref{DD_fm} and \eqref{DD_fp} with $g^{\m\n}$. The explicit torsion terms on the right hand side (RHS) clearly vanish. Furthermore $[\pd_\m^y,D^y_\n]=0$, thus we have
\begin{align}\label{EOM_phipm}
&g^{\m\n}\[\pd_\m^y\pd_\n^y\r+\pd_\m^y D^y_\n\r\]=g^{\m\n}\f_-(D^y_\n D^y_\m\f_+),\nonumber\\
&g^{\m\n}\[\pd_\m^y\pd_\n^y\r-\pd_\m^y D^y_\n\r\]=g^{\m\n}(D^y_\n D^y_\m\f_-)\f_+.
\end{align} 
We now simplify $g^{\m\n}(D^y_\n D^y_\m\f_-)$ as follows (similar expressions can be deduced for $g^{\m\n}(D^y_\n D^y_\m\f_+)$)
\begin{align}\label{DD_f}
&g^{\m\n}(D^y_\m D^y_\n\f_-)
=g^{\m\n} e^{-y\cdot D^y/2}\(D^y_\m+\frac{1}{2}[y\cdot D^y,D^y_\m]\)\nonumber\\
&\times\(D^y_\n+\frac{1}{2}[y\cdot D^y,D^y_\n]\)\f\nonumber\\
\simeq& e^{-y\cdot D^y/2}g^{\m\n}\big[D^y_\m D^y_\n\f+\frac{1}{2}(D^y_\m[y\cdot D^y,D^y_\n]+[y\cdot D^y,D^y_\m]D^y_\n)\f\big]\nonumber\\
\simeq& e^{-y\cdot D^y/2}g^{\m\n}\(D^y_\m D^y_\n\f+y^\r T^\s_{.\r\n}D^y_\m D^y_\s\f\).
\end{align}
The first term can be simplified using the scalar EOM, which is given by
\begin{align}
\frac{1}{\sqrt{-g}}\pd_\m\(\sqrt{-g}g^{\m\n}\pd_\n\)\f+m^2\f=(\square +m^2)\f=0.
\end{align}
Using $\square\f(x)=g^{\m\n}\nabla_\m\nabla_\n\f(x)-T^\m D^y_\m\f(x)=g^{\m\n}D^y_\m D^y_\n\f(x)-T^\m D^y_\m\f(x)$, we have
\begin{align}
g^{\m\n}D^y_\m D^y_\n\f=-m^2\f+T^\m D^y_\m\f.
\end{align}
Combining with the remaining factor, we have
\begin{align}
&\(e^{-y\cdot D^y/2}g^{\m\n}D^y_\m D^y_\n\f\)\f_+\nonumber\\
\simeq&-m^2\r+ \(T^\m D^y_\m\f_-\)\f_+.
\end{align}
The second term in \eqref{DD_f} can be simplified as follows: pulling $e^{-y\cdot D^y/2}$ through horizontal lift operators, we obtain $D^y_\m D^y_\s\f_-$, which is to be evaluated up to $O(1)$. Note that to $O(1)$, we have $\(\pd_\m^y\pm \frac{D^y_\m}{2}\)\f_\mp=0$ from \eqref{partial_phi}, thus
\begin{align}
&(D^y_\m D^y_\s\f_-)\f_+\simeq\big[\(\frac{D^y_\m}{2}-\pd_\m\)\(\frac{D^y_\m}{2}-\pd_\m\)\f_-\big]\f_+\nonumber\\
&=\(\frac{D^y_\m}{2}-\pd_\m\)\(\frac{D^y_\m}{2}-\pd_\m\)\r\nonumber\\
&\simeq \pd_\m^y\pd_\s^y\r.
\end{align}
Collecting all terms, we have the following simplified form of \eqref{EOM_phipm}
\begin{align}
&g^{\m\n}\[\pd_\m^y\pd_\n^y\r+\pd_\m^y D^y_\n\r\]=-m^2\r+\f_-T^\m D^y_\m\f_+ +\nonumber\\
&g^{\m\n}y^\r T^\s_{.\r\n}\pd_\m \pd_\s\r,\nonumber\\
&g^{\m\n}\[\pd_\m^y\pd_\n^y\r-\pd_\m^y D^y_\n\r\]=-m^2\r+\(T^\m D^y_\m\f_-\)\f_+ -\nonumber\\
&g^{\m\n}y^\r T^\s_{.\r\n}\pd_\m \pd_\s\r.
\end{align}
Again we can replace $(D^y_\m\f_-)\f_+\simeq -\pd_\m\r$ and $\f_-D^y_\m\f_+\simeq \pd_\m\r$ in the above. Applying Wigner transform, we obtain the following from sum and difference of the resulting equations
\begin{align}
&(p^2-m^2)W=0,\nonumber\\
&p^\m\(D_\m+ T^\l_{\m\r}p_\l\pd_p^\r\)W=0.
\end{align}
Note that the terms proportional to $T^\m$ cancel. We can see that the dispersion relation remain unchanged. The kinetic equation can be rewritten using explicit form of $D_\m$ as
\begin{align}
p^\m\(\nabla_\m+\G_{\m\n}^\l p_\l\pd_p^\n\)W=0.
\end{align}

\section{Derivation of equation (22)}

Now we derive EOM for components of the Wigner function. Denoting $\tD_\m=D_\m+T^\l_{\m\n}p_\l\pd_p^\n$, the EOM reads
\begin{align}\label{EOM_tD}
\(\g^\m p_\m-m+\frac{i}{2}\g^\m \tD_\m\)W=0.
\end{align}
Plug the decomposition
\begin{align}\label{decomp}
W=S+i\g^5P+V_a\g^a+A_a\g^5\g^a+\frac{1}{2}\s^{ab}T_{ab}.
\end{align}
into \eqref{EOM_tD} and noting that $[\tD_\m,\g^a]=0$, we have
\begin{align}
&(\g^c p_c-m)\big[S+i\g^5P+V_a\g^a+A_a\g^5\g^a+\frac{1}{2}\s^{ab}T_{ab}\big]+\nonumber\\
&\frac{i}{2}\g^c\big[\tD_cS+i\g^5\tD_cP+\g^a\tD_cV_a+\g^5\g^a\tD_cA_a+\frac{1}{2}\s^{ab}\tD_cT_{ab}\big]\nonumber\\
&=0.
\end{align}
Using the following identities of gamma matrices
\begin{align}\label{gamma_product}
&\g^a\g^b=\h^{ab}-i\s^{ab},\quad \g^a\g^b\g^5=\h^{ab}\g^5+\frac{1}{2}\e^{abcd}\s_{cd},\nonumber\\
&\g^a\s^{cd}=i\(\h^{ac}\g^d-\h^{ad}\g^c+i\e^{acdb}\g_b\g^5\),
\end{align}
we can extract the coefficients of the Clifford basis as 
\begin{align}\label{tD}
&1:\(p_a+\frac{i}{2}\tD_a\)V^a-m S,\nonumber\\
&i\g^5:\(p_a+\frac{i}{2}\tD_a\)(iA^a)-mP,\nonumber\\
&\g^d:\(p_d+\frac{i}{2}\tD_d\)S+\(p_a+\frac{i}{2}\tD_a\)iT^{a}{}_{d}-mV_d,\nonumber\\
&\g^b\g^5:\(p_b+\frac{i}{2}\tD_b\)iP+\(p_a+\frac{i}{2}\tD_a\)\(-\frac{T_{cd}}{2}\)\e^{acd}{}_b+mA_b,\nonumber\\
&\s^{ab}:\(p_a+\frac{i}{2}\tD_a\)(-iV_b)-\(p_c+\frac{i}{2}\tD_c\)\frac{A_d}{2}\e^{cd}{}_{ab}-\frac{m}{2}T_{ab}.
\end{align}
Since $W$ is hermitian $\g^{\hat{0}}W^\dagger\g^{\hat{0}}=W$, all components in the decomposition \eqref{decomp} are real, thus we can easily separate \eqref{tD} into real and imaginary parts to have
\begin{align}
&p_aV^a-mS=0,\nonumber\\
&\tD_a V^a=0,\nonumber\\
&p_aA^a=0,\nonumber\\
&\frac{1}{2}\tD_a A^a-m P=0,\nonumber\\
&p^cS-\frac{1}{2}\tD_aT^{ac}-mV^c=0,\nonumber\\
&\frac{1}{2}\tD^cS+p_aT^{ac}=0,\nonumber\\
&p_fP-\frac{1}{4}\tD^aT^{cd}\e_{acdf}=0,\nonumber\\
&\frac{1}{2}\tD^bP+\frac{1}{2}p_aT_{cd}\e^{acdb}+mA^b=0,\nonumber\\
&-p^{[b}V^{f]}+\frac{1}{4}\tD_cA_g\e^{cgdf}=0,\nonumber\\
&\tD^{[d]}V^{f]}+p_cA_g\e^{cgdf}-mT^{df}=0.
\end{align}


\end{document}